\newcommand\apjcls{1}
\newcommand\aastexcls{2}
\newcommand\othercls{3}

\newcommand\papercls{\aastexcls}
\documentclass[tighten, times,twocolumn]{aastex62}  

\if\papercls \apjcls
\usepackage{apjfonts}
\else\if\papercls \othercls
\usepackage{epsfig}
\usepackage{margin}
\usepackage{times}
\fi\fi
\usepackage{ifthen}
\usepackage{natbib}
\usepackage{bm}
\usepackage{amssymb, amsmath}
\usepackage{appendix}
\usepackage{etoolbox}
\usepackage[T1]{fontenc}
\usepackage{paralist}
\usepackage{newtxtext,newtxmath}
\if\papercls \apjcls 
\newcommand\aas{\ref@jnl{AAS Meeting Abstracts}}
\newcommand\dps{\ref@jnl{AAS/DPS Meeting Abstracts}}
\newcommand\maps{\ref@jnl{MAPS}}
\else\if\papercls \othercls
\usepackage{astjnlabbrev-jh}
\fi\fi

\if\papercls \aastexcls
\hypersetup{citecolor=blue, 
            linkcolor=blue, 
            menucolor=blue, 
            urlcolor=blue}  
\else
\usepackage[
bookmarks=true,           
bookmarksnumbered=true,   
colorlinks=true,          
citecolor=blue,           
linkcolor=blue,           
menucolor=blue,           
urlcolor=blue,            
linkbordercolor={0 0 1},  
pdfborder={0 0 1},
frenchlinks=true]{hyperref}
\fi

\if\papercls \othercls

\else

\fi

\providecommand{\adsurl}[1]{\href{#1}{ADS}}

\makeatletter
\patchcmd{\NAT@citex}
  {\@citea\NAT@hyper@{%
     \NAT@nmfmt{\NAT@nm}%
     \hyper@natlinkbreak{\NAT@aysep\NAT@spacechar}{\@citeb\@extra@b@citeb}%
     \NAT@date}}
  {\@citea\NAT@nmfmt{\NAT@nm}%
   \NAT@aysep\NAT@spacechar\NAT@hyper@{\NAT@date}}{}{}

\patchcmd{\NAT@citex}
  {\@citea\NAT@hyper@{%
     \NAT@nmfmt{\NAT@nm}%
     \hyper@natlinkbreak{\NAT@spacechar\NAT@@open\if*#1*\else#1\NAT@spacechar\fi}%
       {\@citeb\@extra@b@citeb}%
     \NAT@date}}
  {\@citea\NAT@nmfmt{\NAT@nm}%
   \NAT@spacechar\NAT@@open\if*#1*\else#1\NAT@spacechar\fi\NAT@hyper@{\NAT@date}}
\makeatother

\makeatletter
\DeclareRobustCommand{\lowcase}[1]{\@lowcase#1\@nil}
\def\@lowcase#1\@nil{\if\relax#1\relax\else\MakeLowercase{#1}\fi}
\makeatother

\DeclareSymbolFont{UPM}{U}{eur}{m}{n}
\DeclareMathSymbol{\umu}{0}{UPM}{"16}
\let\oldumu=\umu
\renewcommand\umu{\ifmmode\oldumu\else\math{\oldumu}\fi}

\if\papercls \othercls

\else

\fi

\let\oldsim=\sim
\renewcommand\sim{\ifmmode\oldsim\else\math{\oldsim}\fi}
\let\oldpm=\pm
\renewcommand\pm{\ifmmode\oldpm\else\math{\oldpm}\fi}
\newcommand\by{\ifmmode\times\else\math{\times}\fi}

\newbox{\wdbox}
\renewcommand\c{\setbox\wdbox=\hbox{,}\hspace{\wd\wdbox}}
\renewcommand\i{\setbox\wdbox=\hbox{i}\hspace{\wd\wdbox}}

\newcount\timect
\newcount\hourct
\newcount\minct
\newcommand\now{\timect=\time \divide\timect by 60
         \hourct=\timect Cltiply\hourct by 60
         \minct=\time \advance\minct by -\hourct
         \number\timect:\ifnum \minct < 10 0\fi\number\minct}

\catcode`@=11

\newcommand\comment[1]{}

\newcommand\commenton{\catcode`\%=14}

\renewcommand\math[1]{$#1$}
\newcommand\mathshifton{\catcode`\$=3}

\let\atab=&
\newcommand\atabon{\catcode`\&=4}

\let\oldmsp=\sp
\let\oldmsb=\sb
\def\sp#1{\ifmmode
           \oldmsp{#1}%
         \else\strut\raise.85ex\hbox{\scriptsize #1}\fi}
\def\sb#1{\ifmmode
           \oldmsb{#1}%
         \else\strut\raise-.54ex\hbox{\scriptsize #1}\fi}
\newbox\@sp
\newbox\@sb
\def\sbp#1#2{\ifmmode%
           \oldmsb{#1}\oldmsp{#2}%
         \else
           \setbox\@sb=\hbox{\sb{#1}}%
           \setbox\@sp=\hbox{\sp{#2}}%
           \rlap{\copy\@sb}\copy\@sp
           \ifdim \wd\@sb >\wd\@sp
             \hskip -\wd\@sp \hskip \wd\@sb
           \fi
        \fi}
\def\msp#1{\ifmmode
           \oldmsp{#1}
         \else \math{\oldmsp{#1}}\fi}
\def\msb#1{\ifmmode
           \oldmsb{#1}
         \else \math{\oldmsb{#1}}\fi}

\def\supon{\catcode`\^=7}

\def\subon{\catcode`\_=8}

\def\supsubon{\supon \subon}

\newcommand\actcharon{\catcode`\~=13}

\newcommand\paramon{\catcode`\#=6}

\comment{And now to turn us totally on and off...}

\newcommand\reservedcharson{ \commenton  \mathshifton  \atabon  \supsubon 
                             \actcharon  \paramon}

\catcode`@=12
\reservedcharson

\if\papercls \apjcls

\else

\fi

\newcommand\chisq{\ifmmode{\chi\sp{2}}\else\math{\chi\sp{2}}\fi}
\newcommand\redchisq{\ifmmode{ \chi\sp{2}\sb{\rm red}}
                    \else\math{\chi\sp{2}\sb{\rm red}}\fi}
\newcommand\Teq{\ifmmode{T\sb{\rm eq}}\else$T$\sb{eq}\fi}
\newcommand\mjup{\ifmmode{M\sb{\rm Jup}}\else$M$\sb{Jup}\fi}
\newcommand\rjup{\ifmmode{R\sb{\rm Jup}}\else$R$\sb{Jup}\fi}
\newcommand\msun{\ifmmode{M\sb{\odot}}\else$M\sb{\odot}$\fi}
\newcommand\rsun{\ifmmode{R\sb{\odot}}\else$R\sb{\odot}$\fi}
\newcommand\mearth{\ifmmode{M\sb{\oplus}}\else$M\sb{\oplus}$\fi}
\newcommand\rearth{\ifmmode{R\sb{\oplus}}\else$R\sb{\oplus}$\fi}

\renewcommand{\bm}[1]{{\mbox{{\boldmath$#1$}}}}	

\newcommand{\grad}{\bm{\nabla}}

\shorttitle{Magnetic field evolution for crystallization-driven dynamos in C\slash O white dwarfs}
\shortauthors{M. Castro-Tapia {\em et al.}}

\begin{document}

\title{Magnetic field evolution for crystallization-driven dynamos in C\slash O white dwarfs}
\author{Matias Castro-Tapia}
\affiliation{\rm Department of Physics and Trottier Space Institute, McGill University, Montreal, QC H3A 2T8, Canada}

\author{Shu Zhang}
\affiliation{\rm Department of Physics and Trottier Space Institute, McGill University, Montreal, QC H3A 2T8, Canada}

\author{Andrew Cumming}
\affiliation{\rm Department of Physics and Trottier Space Institute, McGill University, Montreal, QC H3A 2T8, Canada}


\begin{abstract}
We investigate the evolution of magnetic fields generated by the crystallization-driven dynamo in carbon-oxygen white dwarfs (WDs) with masses $\lesssim1.05\ M_{\odot}$. We use scalings for the dynamo to demonstrate that the initial magnetic field strength ($B_{0}$) has an upper limit that depends on the initial convection zone size ($R_{\mathrm{out},0}$) and the WD mass. 
We solve the induction equation to follow the magnetic field evolution after the dynamo phase ends. We show that the predicted surface magnetic field strength ($B_{\mathrm{surf}}$) differs from $B_{0}$ by at least a factor of $\sim$0.3. This reduction depends on $R_{\mathrm{out},0}$, where values smaller than half of the star radius give $B_{\mathrm{surf}}\lesssim0.01\ B_{0}$. We implement electrical conductivities that account for the solid phase effect on the Ohmic diffusion. We observe that the conductivity increases as the solid core grows, freezing in the magnetic field at a certain point of the evolution and slowing its outwards transport. We study the effect of turbulent magnetic diffusivity induced by the convection and find that for a small $R_{\mathrm{out},0}$, $B_{\mathrm{surf}}$ is stronger than the non-turbulent diffusion cases because of the more rapid transport, but still orders of magnitude smaller than $B_{0}$. Given these limitations, the crystallization-driven dynamo theory could explain only magnetic C/O WDs with field strengths less than a few MG for the mass range 0.45-1.05 $M_{\odot}$. Our results also suggest that a buried fossil field must be at least 100 times stronger than observed surface fields if crystallization-driven convection is responsible for its transport to the surface.
\end{abstract}


\section{Introduction}
The origin of strong magnetic fields in white dwarfs (WDs) remains a challenging problem in astrophysics. The so-called magnetic white dwarfs (MWDs) show magnetic fields stronger than $\sim 10^{4}\ \mathrm{G}$, and up to $\gtrsim10^{9}$ G  has been observed \citep[see][for detailed reviews]{Ferrario2015, Ferrario2020}. Possible explanations for magnetism in WDs include the idea of an emerging fossil field that was formed during previous stages of stellar evolution \citep{Angel1981,BraithwaiteSpruit2004, Tout2004, WickramasingheFerrario2005}, which could imply for example the origin in a main sequence-main sequence merger \citep{Ferrario2009, Schneider2019}, a dynamo generated due to common-envelope evolution \citep{RegosTout1995, Tout2008, PotterTout2010, Nordhaus2011}, and the merger of double-degenerate cores \citep{GarciaBerro2012, Wickramasinghe2014}.

Whereas the common-envelope scenario can explain magnetic cataclysmic variables \citep{Tout2008, PotterTout2010}, the situation for isolated WDs is less clear. A double-degenerate merger or common-envelope evolution leading to a merger can produce strong magnetic fields in ultra-massive WDs, and reproduce some mass and magnetic field distributions of MWDs \citep[e.g.,][]{GarciaBerro2012, Briggs2015, Briggs2018}. However, this scenario predicts that the magnetic field remains in the hot outer layers of the post-merger WD, which means the magnetic field should be detectable shortly after the merger. Therefore, this theory cannot explain why observations show that magnetic WDs preferentially appear at low temperatures and luminosities \citep[e.g.][]{Liebert2003, Sion2014}, which hints at a late emergence of the field and a possible fossil-origin or an alternative channel.

A more recent theory suggests that isolated WDs can sustain a magnetic dynamo following crystallization as a consequence of compositionally-driven convection \citep{Isern2017}. WDs cool over Gyr timescales when they evolve as single stars \citep{Mestel1952}. Due to their high-density degenerate interiors, they experience a phase transition, which crystallizes their core outwards from some point of the cooling \citep{vanHorn1968}. WDs that have been formed with masses $\sim$ 0.5-1 $M_{\odot}$ contain carbon-oxygen cores, which preferentially retain the oxygen in the solid phase when undergoing crystallization, releasing lower-density carbon on top of it \citep[e.g.][]{Isern1997}. This creates a buoyantly unstable region that drives compositional convection as the core continues growing  \citep{Stevenson1980, Mochkovitch1983, Isern1997, Fuentes2023}. This convective motion during crystallization combined with rotation has been proposed to drive a dynamo intense enough to explain some MWDs \citep[e.g.,][]{Isern2017, Ginzburg2022, Fuentes2024}.


However, while less massive MWDs tend to be cooler and older than the expected age to be crystallizing, the more massive ones appear close to or just before the onset of crystallization, which has led to the interpretation that MWDs originate through different channels \citep{BagnuloLandstreet2021, BagnuloLandstreet2022}. Still, for ultra-massive WDs ($M\gtrsim1.05M_{\odot}$), the onset of crystallization is uncertain since it occurs at different times depending on the core composition, C/O or O/Ne  \citep[e.g.][]{Camisassa2019, Camisassa2022a, Camisassa2022b}. Other recent studies have pointed out that the crystallization-driven dynamo can potentially explain the occurrence or not of MWDs in binaries and singles for different stellar populations \citep{Belloni2021, Schreiber2021a, Schreiber2021b, Schreiber2022}, and some specific targets such as the magnetic symbiotic system FN Sgr \citep{Belloni2024}.

Most theoretical works on crystallization-driven dynamos in white dwarfs have focused on studying whether the convection is vigorous enough to drive the dynamo. \citet{Isern2017} and \citet{Ginzburg2022} estimated convective velocities of order $>10^{2}\ \mathrm{cm\ s^{-1}}$, enough to produce MG magnetic fields. They assumed that the kinetic energy associated with convection is about the same as the gravitational energy released by the chemical separation. Recently, however, \citet{Fuentes2023} showed that thermal diffusion, rapid due to the large electron thermal conductivity in WD interiors, can significantly reduce the convective velocity to values $\lesssim 10^{-2}\ \mathrm{cm\ s^{-1}}$. The regime where this occurs corresponds to the regime of thermohaline convection (see also \citealt{MontgomeryDunlap2024, Castro-Tapia2024}). \citet{Castro-Tapia2024} and \citet{Fuentes2024} showed that fast overturning convection can still occur, but only for a few Myr after the onset of crystallization, generating magnetic fields in the range $10^{6}$ to $10^{8}$ G.

A short-lived dynamo in the early stage of crystallization as suggested by \cite{Fuentes2024} produces a magnetic field deep in the interior of the white dwarf. 
In this paper, we calculate the transport of the magnetic field from the interior to the surface following the dynamo phase. Given the importance of this theory to constrain the origin of C/O MWDs with masses $\lesssim 1.05\ M_{\odot}$, we investigate the feasibility of the crystallization-driven dynamo as the only scenario to explain the observed magnetic fields on C/O MWDs with these masses. Recently, \citet{BlatmanGinzburg2024} also investigated this problem for a sample of C/O MWDs showing promising results as the cooling ages of some of the targets roughly coincide with their estimations for the time at which the magnetic field emerges to the surface. However, they only focused on estimating the time at which the magnetic field from the crystallization-driven dynamo would appear on the surface without following its strength evolution as it is transported. We improve on these results by numerically solving the evolution of the magnetic field in the WD interior and calculating the difference from its initial strength as it reaches the surface. This allows us to limit the number of MWDs that can be explained by this theory. We also improve the transport coefficients considered by adding the effect of the solid phase to the magnetic diffusivity and investigating the plausibility of turbulent transport of the field in the region undergoing thermohaline convection.

We start in Section \ref{sec:methods} with a discussion of the fiducial WD cooling models that we use to follow the crystallization, and the relevant stellar parameters used to estimate the initial conditions of the convective motions and the induced magnetic field during the fast convection regime. We combine the MAC (Magnetic-Archimedean-Coriolis) force balance approach of \citet{Fuentes2024} with analytic estimates of the early time evolution after onset of crystallization to restrict the strength of the initial dynamo-generated magnetic field. In Section \ref{sec:B_evol_obs} we present our numerical scheme to solve the evolution of the magnetic field in the WD interior after its formation and compare the resulting surface fields of these transport models with observations of MWDs in the mass range $0.45$--$1.05\ M_{\odot}$. Finally, in Section \ref{sec:summary_conclusions}, we discuss the implications of our results for explaining C/O MWDs with a crystallization-driven dynamo. In particular, we find that only those MWDs with magnetic fields less than about a few MG can be explained with a crystallization-driven dynamo, suggesting that this is not the only formation channel for MWDs in this mass range.

\section{Cooling models and initial magnetic field}\label{sec:methods}

In this section, we first describe the grid of cooling models of carbon-oxygen white dwarfs that we use to compute the relevant parameters expected in their interior during the transport of the magnetic field (Section \ref{sec:cooling_models}). We then calculate the magnetic field strength expected from the crystallization-driven dynamo needed as an initial condition for the transport calculations (Section \ref{sec:initial_magnetic_field}). To calculate the outcome of the dynamo, we follow the MAC balance arguments of \citet{Fuentes2024}, paying careful attention to the changes in the gravity and length scales near the center of the star in the very early stages of crystallization. We show that the dynamo magnetic field reaches a maximum value shortly after crystallization begins (Section \ref{sec:upper_limit}).

\subsection{Cooling models}\label{sec:cooling_models}
We use MESA version r23.05.1 \citep{Paxton2011, Paxton2013, Paxton2015, Paxton2018, Paxton2019, Jermyn2023} to run a grid of WD cooling models from \citet{Bauer2023}. The models have a composition profile obtained from stellar evolution, and masses in the range $0.5$-$1.0\ M_{\odot}$ (with a difference of $\approx 0.1\ M_{\odot}$ between models). We included the effect of phase separation from crystallization \citep{Bauer2023} and used a simplified nuclear network containing only C, O, He, and H (with nuclear reactions turned off). To compute the relevant thermodynamic properties in the stellar interior and the parameters for crystallization-driven convection, we followed the growing solid core using the output from the Skye equation of state \citep{Jermyn2021} as described in \citet{Castro-Tapia2024}. We used the parameter $\phi=0.9$ from Skye ($\phi=0$ is the liquid phase, while $\phi=1$ is the solid) to identify the mass and radius of the solid core at each timestep, and $\phi=0.01$ to evaluate the properties in the liquid zone above the core where the convective instability develops.

In our previous series of papers \citep{Fuentes2023, Fuentes2024, Castro-Tapia2024}, we have shown that the composition flux of carbon $F_{X}$ into the C/O liquid mixture out of the solid core determines the efficiency of convection at each stage of the cooling. We also defined a dimensionless parameter to describe this compositional flux,
\begin{equation}\label{eq_tau}
    \tau=F_{X}\frac{H_{p}\chi_{X}}{\rho\kappa_{T}\chi_{T}X\nabla_{\mathrm{ad}}}~,
\end{equation}
where $\nabla_{\rm ad}$ is the adiabatic temperature gradient, $\chi_T = \left.\partial\ln P/\partial\ln T\right|_{\rho, X}$, $\chi_X = \left.\partial\ln P/\partial\ln X\right|_{\rho, T}$, $\kappa_{T}=4acT^{3}/(3\kappa \rho^{2}c_{P})$ is the thermal diffusivity, where $\kappa$ is the opacity, $a$ the radiation constant, $c$ the speed of light, and $c_{P}$ is the specific heat capacity at constant pressure. This expression for $\tau$ can be rewritten in terms of the thermal timescale $t_{\mathrm{therm}}=H_{P}^{2}/\kappa_{T}$ (across the pressure scale height $H_{P}$) and the characteristic timescale for the injection of light elements $t_X=\rho H_P X/F_X$ as $\tau=(t_\mathrm{therm}/t_X)(\chi_X/\chi_T\nabla_{\mathrm{ad}})$. Using the mass and radius of the crystallized core we can write the flux $F_{X}$ as follows
\begin{equation}\label{eq_Fx}
    F_X = \frac{\dot{M}_{\mathrm{core}}\Delta X}{4\pi R^2_{\mathrm{core}}},
\end{equation}
where $M_{\mathrm{core}}$ and $R_{\mathrm{core}}$ are the mass and radius of the solid core at a given time, and $\Delta X$ is the mass fraction enhancement of carbon in the liquid region relative to the solid when fractionation occurs. $\Delta{X}$ is computed following the analytical fitting of the C/O phase diagram for giving local O abundance presented by \citet{Blouin2020, Blouin2021b}. 

\citet{Fuentes2024} derived an analytical expression for $R_{\mathrm{core}}$ in the early stages of crystallization that depends only on the cooling history of the white dwarf. Solving hydrostatic equilibrium for an almost constant central density $\rho_{c}$, using the equation of state for degenerate electrons $P=K\rho^{5/3}$, and assuming that crystallization starts when the coupling parameter $\Gamma_{c}\propto \rho^{1/3}/T$ (the ratio of the electrostatic to thermal energy) at the center of the star exceeds the almost constant critical value $\Gamma_{c,\mathrm{crit}} \approx 175$ \citep{vanHorn1968, Potekhin2000, Bauer2020, Jermyn2021} gives
\begin{equation}
    \dfrac{R_{\mathrm{core}}}{R_{\rm WD}} = \left[1 - \left(\dfrac{T_{\mathrm{core}}}{T_i}\right)^2\right]^{1/2}~, \label{eq:Rcore/R}
\end{equation}
where $R_{\mathrm{WD}}$ is the radius of the white dwarf, and $T_{\mathrm{core}}$ is the temperature of the core for a given time, and $T_i=T_{\mathrm{core}}(t=t_{\mathrm{crys}})$, with $t_{\mathrm{crys}}$ the time at which crystallization starts. Then for an incompressible core $M_{\mathrm{core}} = (4\pi/3) \rho_c R^3_{\mathrm{core}}$, we write $\dot{M}_{\mathrm{core}} = 4\pi \rho_c R^2_{\mathrm{core}}\dot{R}_{\mathrm{core}}$, and combining equations \eqref{eq_tau}, \eqref{eq_Fx}, and \eqref{eq:Rcore/R} gives
\begin{equation} \label{eq:tau_analytic}
\tau = \alpha \left(\dfrac{\dot{T}_{\mathrm{core}}}{T_i}\right)  \dfrac{(T_i/T_{\mathrm{core}})^3}{\sqrt{1-(T_{\mathrm{core}}/T_i)^2}}~,
\end{equation}
where $\alpha = -R_{\mathrm{WD}}(H_P/\kappa_T)(\Delta X/X)(\chi_X/\chi_T\nabla_{\mathrm{ad}})$. 

\begin{figure*}
    \centering
    \includegraphics[width=0.97\textwidth]{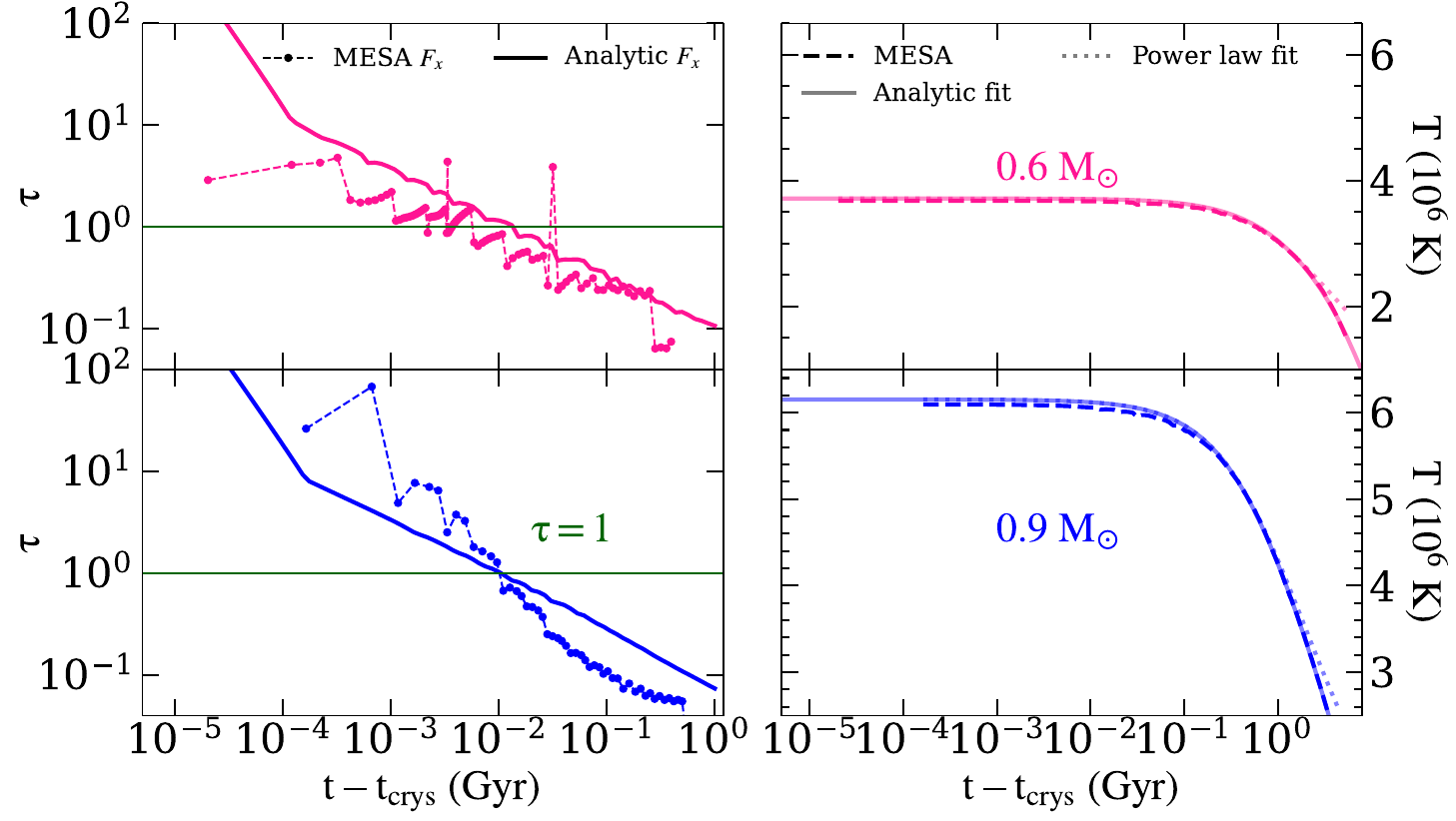} 
  \caption{\textit{Left panels}: Normalized composition flux $\tau$ computed for the MESA cooling WD models of $0.6$ (pink) and $0.9\ M_{\odot}$ (blue) using the estimation of $F_{X}$ directly from the MESA output of the mass $M_{\mathrm{core}}$ and radius $R_{\mathrm{core}}$ of the crystallized core (dashed lines with data points), and using the analytic fit shown in Eq.~\eqref{eq:tau_analytic} (solid lines). \textit{Right panels}: Cooling tracks for the core temperature of the MESA WD models (dashed lines) along with the analytic fit done from Eq.~\eqref{eq_T_analytic} (solid lines, and dotted for only the power-law component) and used to compute the analytical fit for $\tau$. The bumps in the analytic curves are due to interpolation of the stellar parameters used to compute $\tau$.}
    \label{fig:cooling_tau}
\end{figure*}

In the left panels of Figure \ref{fig:cooling_tau} we show a comparison of $\tau$ obtained by computing $F_{x}$ directly from the cooling models and by using the analytic expression \eqref{eq:tau_analytic} for two different masses. To obtain $T_{\mathrm{core}}$ at each time we fit an analytical formula to the cooling curves from the MESA models as shown in the right panels of Figure \ref{fig:cooling_tau}. This analytic fit depended on two different parameters $n$ and $\sigma$ for each mass

\begin{equation}\label{eq_T_analytic}
    T_{\mathrm{core}}=T_{i}\left(\frac{t}{t_{\mathrm{crys}}}\right)^{n}e^{-(t/t_{\mathrm{crys}}-1)^{2}/\sigma^{2}}.
\end{equation}
We also note that $\alpha$ in equation \eqref{eq:tau_analytic} slightly changes over time, then we took the corresponding local values at each time for each stellar mass and interpolated to make the analytic fits for $\tau$. Having an analytical fit for the core temperature allows us to predict $\tau$ at times very close to the onset of crystallization. The functional form chosen to do the analytic fit for the core temperature shown in eq.~\eqref{eq_T_analytic} is motivated by the power law derived by \citet{Mestel1952}. However, we see that the cooling rate in our WD models at ages of about $\gtrsim1$ Gyr from the onset of crystallization is faster than predicted by a power law fit component only. Thus, we included a Gaussian component that allows an enhanced decrease of $T_{\mathrm{core}}$, better fitting the cooling tracks.
We show the values obtained for $n$ and $\sigma$ in Table \ref{tab:tabpar} for each MESA model. Note that the power law coefficient $n$ for each model is $\sim -0.5$, which is similar to the Mestel cooling law $T_{\mathrm{core}}\propto t^{-2/5}$.

\begin{table}
	\centering
	\caption{Coefficients obtained for the analytic fit of the core temperature as a function of time as shown in eq.~\eqref{eq_T_analytic}. The $n$ coefficient is for the power law component and $\sigma$ for the Gaussian.}
	\label{tab:tabpar}
	\begin{tabular}{ccc} 
		\hline
	Mass ($M_{\odot}$)  &$n$&$\sigma$\\
    		\hline
        0.5 &-0.49&3.45\\
        0.6 &-0.45&5.29\\
        0.7 &-0.46&7.62\\
        0.8 &-0.51&9.94\\
        0.9 &-0.50&10.16\\
        1.0 &-0.52&10.43\\
		\hline
  
	\end{tabular}
\end{table}

The change from $\tau>1$ to $\tau<1$ marks the transition between efficient and inefficient convection \citep{Fuentes2023, Castro-Tapia2024}, with the efficient case being where strong magnetic fields ($\gtrsim10^{5}\ \mathrm{G}$) can be induced \citep{Fuentes2024}. We see that the analytic formula for $\tau$ in Figure \ref{fig:cooling_tau} shows the transition between convection regimes at close to the same time as the numerical values from MESA. In the following section, we use equations \eqref{eq:tau_analytic} and \eqref{eq_T_analytic} to compute the relevant convective velocities associated with the composition fluxes at each time.

\subsection{Convective velocities and initial magnetic field}\label{sec:initial_magnetic_field}
The relevant parameters associated with convection mainly depend on the density contrast $D\rho$ that drives the buoyant instability. Including the effect of thermal  and composition contrasts $DT$ and $DX$ (for a two-component plasma with abundances $X$ and $Y=1-X$) we can write

\begin{align}\label{eq_del_rho_comp}
    \nonumber\frac{D\rho}{\rho}&=-\frac{\chi_{T}}{\chi_{\rho}}\frac{DT}{T}-\frac{\chi_{X}}{\chi_{\rho}}\frac{DX_{i}}{X}\\
    &=-\frac{\chi_{T}}{\chi_{\rho}}(\nabla-\nabla_{e}+\nabla_{\mathrm{com}})\frac{\ell}{2H_{P}}~, 
\end{align}
where $\chi_\rho = \left.\partial\ln P/\partial\ln \rho\right|_{T, X}$, $\ell$ is the mixing length, $\nabla=d\ln{T}/d\ln{P}|_{\star}$ is the temperature gradient with pressure in the star, $\nabla_{e}$ is the temperature gradient with pressure experienced by a fluid element rising due to the buoyant instability, and $\nabla_{\mathrm{com}}=(\chi_{X}/\chi_{T})\nabla_{X}$ is the local composition gradient, where $\nabla_{X}=d \ln X/d \ln P|_{\star}$. The buoyant acceleration is proportional to $-D\rho/\rho$ and therefore the condition for continued convection is $D\rho<0$ or

\begin{equation}\label{eq_continuous_conv}
\nabla-\nabla_{e}+\nabla_{\mathrm{com}}>0.
\end{equation}

In \citet{Castro-Tapia2024}, we used the composition flux $\tau$ and the total heat flux (represented by the temperature gradient needed to transport all the energy by radiation $\nabla_{\mathrm{rad}}$) to solve for the convective velocity and the relevant thermal and composition gradients presented in equation \eqref{eq_del_rho_comp}. The total heat flux $F=\rho c_{P} \kappa_{T}(T/H_{P})\nabla_{\mathrm{rad}}$ is given by the summation of the heat transported by radiation $F_{\mathrm{rad}}=\rho c_{P} \kappa_{T}(T/H_{P})\nabla$ and the heat transported by convection $ F_c = \rho v_c c_{P} DT = \rho v_c c_{P} T\left(\nabla-\nabla_e\right)(\ell/2H_{P})$, where $v_{c}$ is the convective velocity. Then the difference in the thermal gradients from the convective heat flux can be written as

\begin{equation}\label{del1}
     \left(\nabla-\nabla_e\right)=\frac{2\kappa_{T}}{v_{c}\ell}(\nabla_{\mathrm{rad}}-\nabla),
\end{equation}
and using the definition of the convective efficiency from eq. (3) of \citet{Castro-Tapia2024}

\begin{equation}\label{gamma}
    \Gamma =  \frac{\nabla-\nabla_{e}}{\nabla_{e}-\nabla_{\mathrm{ad}}} = \frac{1}{2a_{0}}\frac{v_{c}\ell}{\kappa_{T}},
\end{equation}
where $a_{0}$ is a constant shape parameter of order of unity, eq.~\eqref{del1} can be written as

\begin{equation}\label{del_del_e}
    \left(\nabla-\nabla_e\right)=\frac{\Gamma(\nabla_{\mathrm{rad}}-\nabla_{\mathrm{ad}})}{1+\Gamma(1+a_{0}\Gamma)}.
\end{equation} 
Thus, the difference in the thermal gradients in eq.~\eqref{eq_del_rho_comp} only depends on the value of $\Gamma$ (or the Peclet number $\mathrm{Pe}$) for given $\nabla_{\mathrm{rad}}$ and $\nabla_{\mathrm{ad}}$.

We have described in the previous section how to write the composition flux in terms of the amount of carbon expelled into the liquid phase as the core crystallizes. However, in general, the composition flux can be written in terms of the specific composition gradient of $X$ as $F_{X}=\rho v_{c}DX=\rho v_{c}X\nabla_{X}(\ell/2H_{P})$. Using equations \eqref{eq_tau} and \eqref{gamma}, we can relate the composition gradient to $\tau$ and the convective efficiency:
\begin{equation}\label{com_tau}
\nabla_{\mathrm{com}}=\frac{\tau\nabla_{\mathrm{ad}}}{a_{0}\Gamma}.
\end{equation}
This is useful because it allows the composition gradient in equation \eqref{eq_del_rho_comp} to be rewritten in terms of $\tau$.

Now, using the balance between the magnetic force, buoyancy, and the Coriolis force (MAC balance; \citealt{Davidson2013}) as presented in \citet{Fuentes2024} we write the buoyant term $-(g/ \ell_{\perp})(D\rho/\rho)$ replacing the gradients in eq.~\eqref{eq_del_rho_comp} by equations \eqref{del_del_e} and \eqref{com_tau} as follows  (assuming $\ell\approx\ell_{\perp}$)

\begin{equation}\label{eq:MAC}
  \frac{u^2_A} {\ell_{\perp}^2}\approx {g\over 2H_{P}}\frac{\chi_{T}}{\chi_{\rho}}\left[\frac{\Gamma(\nabla_{\mathrm{rad}}-\nabla_{\mathrm{ad}})}{1+\Gamma(1+a_{0}\Gamma)}+\frac{\tau\nabla_{\mathrm{ad}}}{a_{0}\Gamma}\right]\approx {2\Omega v_c\over H_P}, 
\end{equation}
where $u_A \equiv B/\sqrt{4\pi \rho}$ is the Alfv\'en speed, with $B$ the strength of the magnetic field, and $\Omega$ is the angular rotation rate. Furthermore, from equation (20) of \citet{Fuentes2024} we have an expression for the mixing length depending on $v_{c}$ and $u_{A}$ separately
\begin{equation}\label{eq:L}
\dfrac{\ell}{H_P} \approx \left(\dfrac{v_c}{2\Omega H_P}\right)^{1/4} \approx \left(\dfrac{u_A}{2\Omega H_P}\right)^{1/3},
\end{equation}
which is obtained by combining the MAC balance and the scaling law derived from the argument of \citet{Davidson2013} that for flows that have small values of the magnetic Prandtl number, $\mathrm{Pm}$ ($\sim0.5$ in C/O WDs), the rate of work done by the buoyancy force determines the magnetic energy density. 

Thus, the two first terms in eq.~\eqref{eq:L} combined with the first and last term of \eqref{gamma} give an expression of the convective efficiency $\Gamma$ in terms of $v_{c}$, $H_{P}$, and $\Omega$

\begin{equation}\label{gamma_vc}
    \Gamma=\frac{1}{2a_{0}}\frac{v_{c}^{5/4}H_{P}^{3/4}}{\kappa_{T}(2\Omega)^{1/4}}.
\end{equation}
And finally, the two last terms of eq.~\eqref{eq:MAC} combined with eq.~\eqref{gamma_vc} gives an equation to solve for $v_{c}$ (or $\Gamma$) for given stellar parameters, fluxes $\tau$ and $\nabla_{\mathrm{rad}}$, and rotation rate $\Omega$. From the solution for $v_{c}$, we can write an expression for the magnetic field using eq.~\eqref{eq:L}
\begin{equation}\label{eq:B}
    B\approx v_{c}^{3/4}(2\Omega H_{P})^{1/4}(4\pi\rho)^{1/2}.
\end{equation}
This generalizes the results of \citet{Fuentes2024} so that we can use a solution for $v_{c}$ and compute $B$ without making any assumption about the relevant gradients for each convective efficiency regime. 

Knowing the cooling history of the WD and the local values of the thermodynamic properties at each stage of crystallization, we computed $\tau$ using the analytical formula \eqref{eq:tau_analytic}, and then $v_{c}$ and $B$ using equations \eqref{eq:MAC}, \eqref{gamma_vc}, and \eqref{eq:B} for each model. Since most of the thermodynamic properties remain almost constant near the center of the WD in the early crystallization phase, we fixed those parameters to be the central values given in the models at the defined $t_{\mathrm{crys}}$ in each case. 
Since at $t_{\mathrm{crys}}$ no solid has been formed yet, $R_{\mathrm{core}}=0$, which implies a divergence in the value of $\tau$. However, for an almost constant central density $\rho_{c}$, the local gravity $g\propto\rho_{c} R_{\mathrm{core}}$ goes to $0$ when no solid is formed, which means that the middle term in eq.~\eqref{eq:MAC} vanishes, giving $v_{c}=0$ (and $B=0$).

The pressure scale height $H_{p}=-P(dP/dr)^{-1}$ also diverges towards the center for hydrostatic equilibrium $dP/dr=-\rho g$ and zero gravity. However, while $H_{P}$ is a local quantity, in mixing length theory we must understand it as a characteristic length over which the temperature and composition gradients develop. Therefore, we are limited by the size of the convection zone, and we replace $H_{P}$ in our equations with $H=\min\{H_{P}, R_{\mathrm{out}}-R_{\mathrm{core}}\}$, where $R_{\mathrm{out}}$ is the outermost radial coordinate of the convection zone. Another key quantity is the mixing length $\ell$, which is naturally limited by $H$ in eq.~\eqref{eq:L}, and since in a rotating fluid, the convective motions are anisotropic, $\ell$ is used to describe the lengthscale of convective motions perpendicular to the rotation axis, whereas $H$ is parallel to $\Omega$ \citep[see][]{Aurnou2020}.  

In order to compute the gravity in the initial stage of crystallization between $t_{\mathrm{crys}}$ and the first time step that MESA outputs show crystallization, we write $g=f_{r}(4\pi/3)G\rho_{c}R_{\mathrm{core}}$. Here, $f_{r}=(1+\Delta{r}/R_{\mathrm{core}})$ is a numerical factor that approximately accounts for the radius difference $\Delta{r}$ between the solid-liquid boundary and the liquid layer where we evaluate the properties of convection. From our MESA models, we see that, for the earlier times after crystallization, $R_{\mathrm{core}}$ at $\phi=0.9$ is about 100 times smaller than the radius coordinate at $\phi=0.01$ where we set to be the unstable liquid phase, which means a difference of a factor of 100 in the gravity. However, since $B\propto g^{1/3}$ for $\tau>1$ (see section \ref{sec:upper_limit}), the difference in $B$ is up to a factor of $f_{r}^{1/3}\sim 4.6$ when considering the gravity in the liquid or the solid. Thus, for all our cooling models we computed $f_{r}$ considering the ratio of the radial coordinate at $\phi=0.01$ and $\phi=0.9$ in the first time step registered after $t_{\mathrm{crys}}$.

\begin{figure*}
    \centering
    \includegraphics[width=0.97\textwidth]{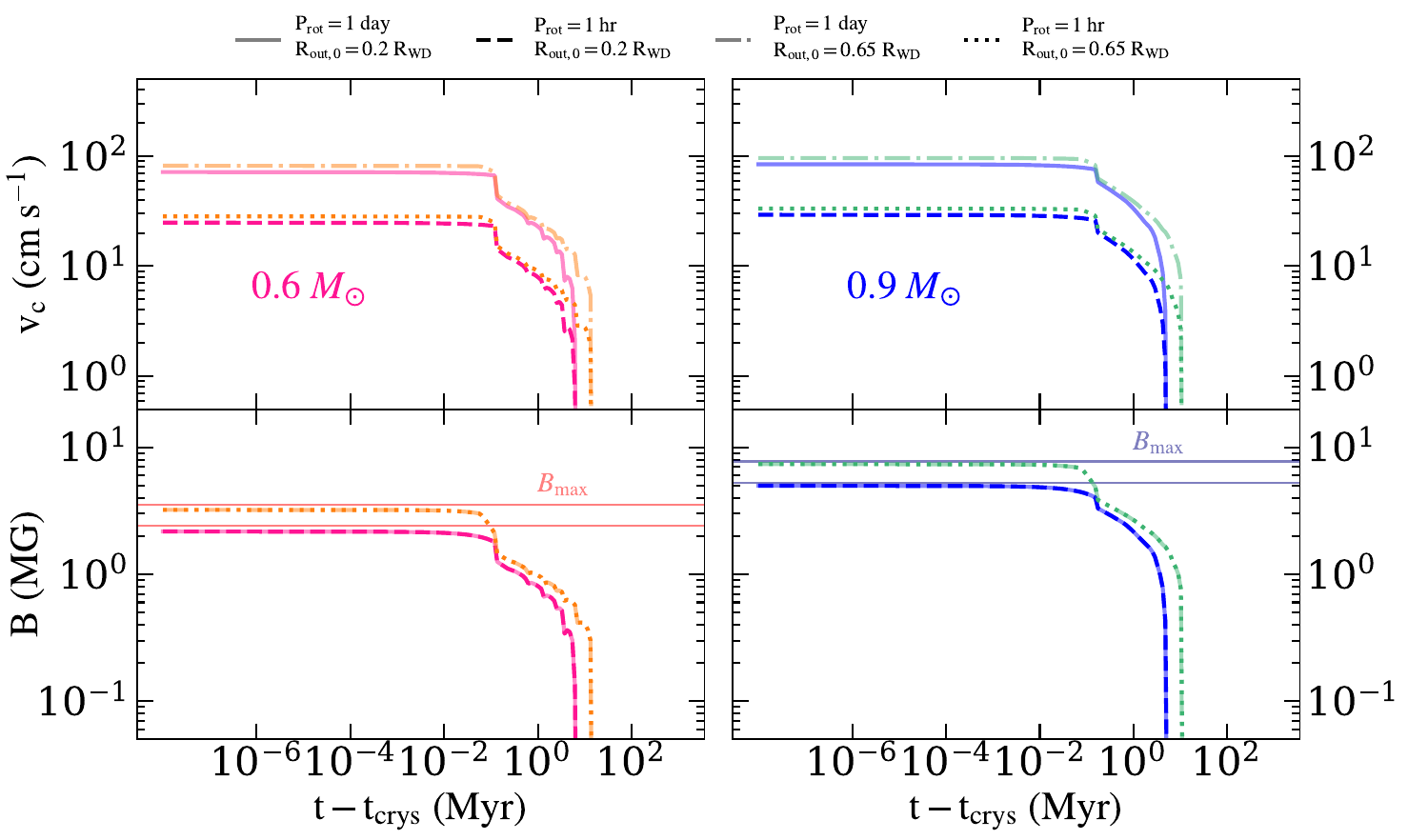} 
  \caption{\textit{Top panels}: Convective velocities $v_{c}$ estimated for the MESA cooling WD models of $0.6$ (pink and orange) and $0.9\ M_{\odot}$ (blue and green), obtained from the numerical solution of equations for the MAC balance \eqref{eq:MAC} and the relation between $v_{c}$ and the convective efficiency $\Gamma$ \eqref{gamma_vc}. \textit{Bottom panels}: Magnetic field strength derived from the convective velocities obtained in the top panels as shown by eq.~\eqref{eq:B}. For both sets of panels, we show the cases for different assumptions of the rotation period $\mathrm{P_{rot}}$ and the initial size of the convection zone $R_\mathrm{out,0}$ (see legends for more details).}
    \label{fig:B_vc}
\end{figure*}

In Figure \ref{fig:B_vc}, we show the predicted values of $v_{c}$ and $B$ as a function of time from crystallization. We show results for masses 0.6 and 0.9 $M_{\odot}$, and for two different assumed rotational periods $\mathrm{P_{rot}}$, and two values of the initial outer radial coordinate of the convection zone $R_\mathrm{out,0}$. We leave $R_\mathrm{out,0}$ as a free parameter to account for the uncertainty in the initial location of the convective boundary\footnote{Figure 1 of \citet{Fuentes2024} shows that the extent of the convection zone depends on the assumptions made for mixing in the liquid region above the solid core.}. We see that the convective velocity is affected by rotation and the extent of $R_\mathrm{out,0}$, whereas the magnetic field does not depend on the rotation rate as pointed out by \citet{Fuentes2024}. Note that we are only showing the predictions of $v_{c}$ and $B$ for the efficient regime since after about 10 Myr both quantities decrease by several orders of magnitude as predicted in \citet{Castro-Tapia2024} and \citet{Fuentes2024}. 

\subsection{An upper limit for the dynamo-generated field}\label{sec:upper_limit}

Figure \ref{fig:B_vc} shows that $v_{c}$ and $B$ are approximately constant for early enough times following crystallization. This is important because we can use this maximum value as an upper limit for the dynamo-generated field. To see why $B$ is constant despite the fact that $\tau$ diverges at early times, we can use the expression for $B$ obtained by \citet{Fuentes2024} for the efficient convection regime ($\tau>1$),
\begin{align}\label{eq:B2}
\nonumber B \approx 1.4~\mathrm{MG}~\left(\dfrac{\tau}{10}\right)^{1/3} &\left(\dfrac{\rho}{10^7~\mathrm{g~cm^{-3}}}\right)^{1/2} \left(\dfrac{g}{10^8~\mathrm{cm~s^{-2}}}\right)^{1/3}\\
&\times\left(\dfrac{\kappa_T}{10~\mathrm{cm^2~s^{-1}}}\right)^{1/3} \left(\dfrac{\chi_T\nabla_{\mathrm{ad}}/\chi_{\rho}}{10^{-4}}\right)^{1/3}~.
\end{align}
When the core is small we have $g\propto R_{\mathrm{core}}\propto \sqrt{1-(T_\mathrm{core}/T_i)^2}$ (eq.~\ref{eq:Rcore/R}), which cancels the $1/\sqrt{1-(T_\mathrm{core}/T_i)^2}$ divergence in $\tau$ (eq.~\ref{eq:tau_analytic}) at early times when $T_\mathrm{core}\approx T_i$. In addition, towards the center $\rho$, $\kappa_{T}$, and $\chi_T\nabla_{\mathrm{ad}}/\chi_{\rho}$ remain almost constant, which means that  $B\propto (g\tau)^{1/3}$ is only weakly-dependent on time.

Assuming $n=-1/2$, $\Delta X=0.1$, and that the cooling is dominated only by the power law component at the early times, we obtain the asymptotic value of $B$ for the efficient regime (combining the expression for the gravity and equations \eqref{eq:Rcore/R}, \eqref{eq:tau_analytic}, and \eqref{eq:B2})\footnote{We have multiplied a factor of 5/6 to eq.~\eqref{eq:B2} since \citet{Fuentes2024} estimated that for the efficient regime $\nabla_{\mathrm{com}}\approx 6(\nabla_{e}-\nabla)$. However, they used this argument to ignore $(\nabla_{e}-\nabla)$ in their approximation, which gives a result off by a factor of about 5/6.}
as
\begin{align}\label{eq:Bmax}
\nonumber B_{\mathrm{max}} \approx 0.89~\mathrm{MG}~\left(\frac{H}{10^{8}\ \mathrm{cm}}\right)^{1/3} \left(\frac{\rho}{10^7~\mathrm{g~cm^{-3}}}\right)^{5/6} \left(\frac{R_{\mathrm{WD}}}{10^8~\mathrm{cm}}\right)^{2/3}\\
\times\left(\dfrac{t_{\mathrm{crys}}}{\mathrm{Gyr}}\right)^{-1/3}
\left(\dfrac{\chi_X/X\chi_{\rho}}{10^{-3}}\right)^{1/3} \left(\frac{f_{r}}{100}\right)^{1/3}.
\end{align}
We added this estimation of the asymptotic value of $B$ to Figure \ref{fig:B_vc} for each case (horizontal lines). Interestingly, the expression in eq.~\eqref{eq:Bmax} is exactly time-independent for the choice $n=-1/2$; for power law cooling the scaling is $B\propto t^{-(2n+1)/3}$ which becomes time-independent for $n=-1/2$.

\section{Magnetic field evolution and comparison to observations}\label{sec:B_evol_obs}

In this section, we describe the numerical scheme used to evolve the internal magnetic field generated through the dynamo theory and compare the surface value obtained with observed data on the strength of magnetic fields in WDs. 

\subsection{Numerical method}

Because the efficient stage of convection in which the dynamo operates lasts for only a short time following crystallization, for the evolution calculation we start with the assumption that the magnetic field is already formed and evolves only by the effect of diffusion. Then we write the induction equation as follows
\begin{equation}\label{eq_ind}
    \frac{\partial\bm{B}}{\partial{t}}=-\grad\times(\eta\grad\times\bm{B}),
\end{equation}
where $\eta=\eta_{\mathrm{ohm}}+\eta_{\mathrm{turb}}$ is the magnetic diffusivity, that we have written as the summation of the Ohmic diffusivity $\eta_{\mathrm{ohm}}=c^{2}/4\pi\sigma$, where $\sigma$ is the electrical conductivity, and a turbulent diffusivity component $\eta_{\mathrm{turb}}=\mathrm{f_{Rm}}\eta_{\mathrm{ohm}}$, with $\mathrm{f_{Rm}}$ proportional to the magnetic Reynolds number $\mathrm{Rm}=v_{c}\ell/\eta_{\mathrm{ohm}}$ \citep[e.g.][]{Denisov2008, TobiasCattaneo2013}.

\begin{figure*}
    \centering
    \includegraphics[width=0.61\textwidth]{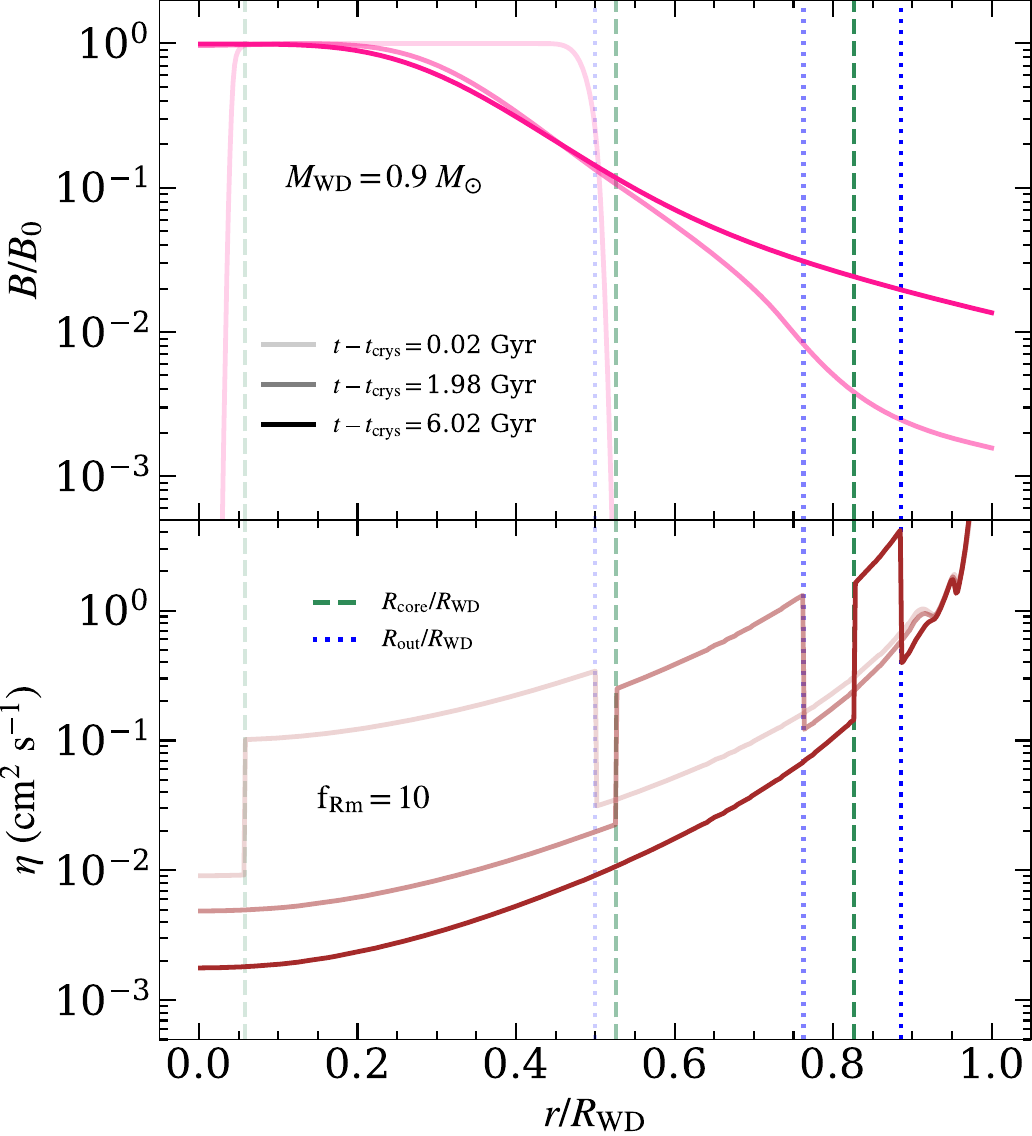}   \includegraphics[width=0.252\textwidth]{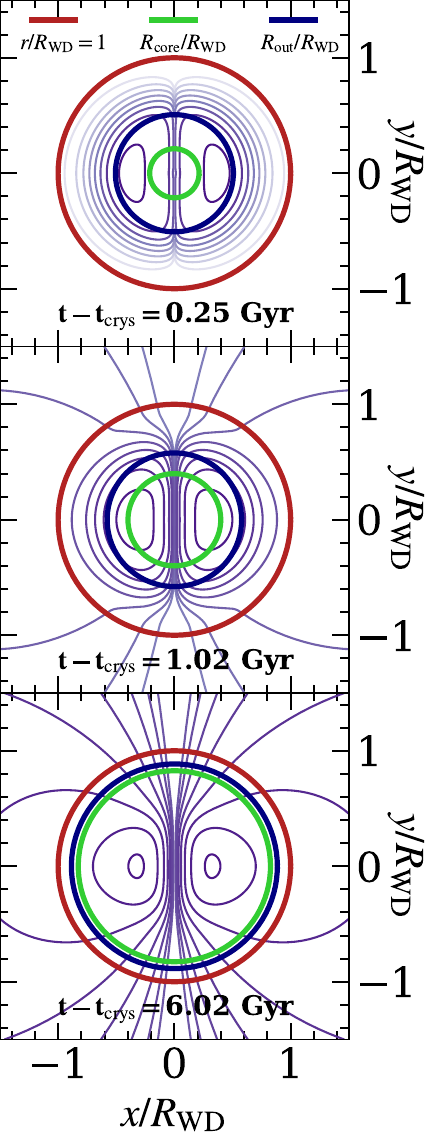}
  \caption{\textit{Left panels}: Evolution of the magnetic field strength and the magnetic diffusivity as a function of the WD radius for the $0.9\ M_{\odot}$ model. We numerically solved the induction equation for the radial component (eq.~\eqref{eq_Rl}) and assumed a dipole geometry to compute the strength of the field using eq.~\eqref{eq_B_vec}. \textit{Right panels}: Two-dimensional visualization of the evolution of the magnetic field lines with equal magnetic flux for the magnetic field obtained for the $0.9\ M_{\odot}$ model (as in the left panels) using eq.~\eqref{eq_B_vec}, the definition of the magnetic vector potential, and assuming and axisymmetric dipole field.}
    \label{fig:B_evol}
\end{figure*}

We used the Electron Transport Coefficients of Magnetized Stellar Plasmas code of \citet{Potekhin1999, Potekhin2015} to compute the electrical conductivity in the WD interior. We took the temperature, density, average charge, and atomic mass of each radial coordinate and timestep for our models, and constructed a time-dependent profile of conductivities. In the solid phase, the electron collision frequency that enters into the conductivity is dominated by phonon scattering and the impurities of the crystal, whereas in the liquid, electron-ion scattering sets the collision frequency. Thus, we write $\sigma\equiv\sigma(r,t)$ in the diffusivity factor for the induction equation. The value of turbulent diffusivity depends on the properties of convection. Once convection enters the inefficient regime ($\tau <1$) and the dynamo turns off (after a few Myr), convection occurs in the thermohaline regime for which we estimate $\mathrm{Rm}\sim10$-$10^{4}$. We let $\mathrm{f_{Rm}}$ be a free parameter varying in such a range of values.

Following earlier work on ohmic diffusion in white dwarfs \citep{ChanmuganGabriel1972, Fontaine1973, Wendell1987,Cumming2002}, we assume an axisymmetric poloidal field, with a magnetic vector potential $\bm{A}=A_{\phi}(r,\theta,t)\hat{e}_{\phi}=\sum_{l}(R_{l}(r,t)/r)P^{1}_{l}(\cos{\theta})\hat{e}_{\phi}$, where we have separated the radial and angular parts, and $P^{1}_{l}$ is the associated Legendre polynomial of order 1. Then using $\bm{B}=\grad\times\bm{A}$ we can rewrite eq.~\eqref{eq_ind} in terms of the magnetic vector potential, and since we assumed spherically symmetric conductivities, the diffusion equation for the radial components $R_{l}$ is

\begin{equation}\label{eq_Rl}
    \frac{\partial R_{l}}{\partial t}=\frac{c^{2}(1+\mathrm{f_{Rm}})}{4\pi\sigma(r,t)}\left[\frac{\partial^{2} R_{l}}{\partial r^{2}}-\frac{l(l+1)R_{l}}{r^{2}}\right].
\end{equation}

During the efficient convection regime where the dynamo acts to generate a strong magnetic field, the size of the solid core is several orders of magnitude smaller than the size of the convection zone $R_{\mathrm{core}}/R_{\mathrm{out}}\lesssim 10^{-4}$. Under this condition, we assume that the dipole component of the field dominates and sets $\ell=1$. Therefore, the components of the magnetic field are
\begin{equation}\label{eq_B_vec}
    \bm{B}(r,\theta,t)=\frac{2R_{1}(r,t)}{r^{2}}\cos{\theta}\hat{e}_{r}+\frac{1}{r}\frac{\partial R_{1}(r,t)}{\partial{r}}\sin{\theta}\hat{e}_{\theta}.
\end{equation}
Taking the same boundary conditions as \citet{Cumming2002}, motivated by the discussion of \citet{Mestel1999}, we fixed $\partial R_{1}/\partial r=2R_{1}/r$ for the center of the star $r\rightarrow{0}$ and  $\partial R_{1}/ \partial r=-R_{1}/r$ for $r\rightarrow{R_{WD}}$ which enforces vanishing current $\bm{J}\propto\grad\times \bm{B}=0$ outside the star. 

We assume that the dynamo-generated field occupies the whole convection zone initially, with \begin{equation}
    |\bm{B}|(r,t_{0})={2R_{1}(r,t_{0})\over r^{2}}=
    \begin{cases} 
      B_{0} & R_{\mathrm{core}}\leq r \leq R_{\mathrm{out,0}}\\
     0 & \mathrm{otherwise}
   \end{cases}.
\end{equation}
In order to be as optimistic as possible in predicting the surface magnetic field strength, we take the initial field $B_{0}$ to be the maximum dynamo-generated field derived in Section \ref{sec:upper_limit}. 

We solve equation \eqref{eq_Rl} using the Crank–Nicolson method to obtain the internal evolution of the initial magnetic field over time for different WD masses and the parameters $R_{\mathrm{out,0}}$ and $\mathrm{f_{Rm}}$. 

\subsection{Example evolution}

Figure \ref{fig:B_evol} shows an example evolution. The left panels show the evolution of the magnetic field of a 0.9 $M_{\odot}$ WD model along with the evolution of the internal diffusivity. In this case, we took the initial outermost coordinate of the convection zone to be $R_{\mathrm{out,0}}=0.5R_{\mathrm{WD}}$ and the turbulent diffusivity $\mathrm{f_{Rm}}=10$. We set the initial time $t_{0}-t_{\mathrm{crys}}=0.01$ Gyr corresponding to the time when $\tau$ crosses unity, at which point convection transitions from the efficient regime to the thermohaline regime and the dynamo ends. We then evolve in time, and the initial $B_{0}$ diffuses outwards. Because of the growth of the solid core, the diffusivity $\eta\propto\sigma^{-1}$ gradually decreases in the interior as the conductivity increases over time. At the later times ($2$ and $6\ \mathrm{Gyr}$), the magnetic field at the surface grows to $\approx 1$\% of the initial dynamo field $B_0$.

In the right panels of Figure \ref{fig:B_evol} we show the magnetic field lines with the same magnetic flux value in each panel for different times (given our definition of $A_\phi$, the quantity $A_\phi r\sin\theta$ is constant along magnetic field lines). While we do not show units for the lines, the intensity of the field is proportional to the opacity of the line colors. We also included the evolution of the radius of the solid core and the convection zone. This visualization of the result shows the same behavior as the upper left panel of the figure where the intensity of the field is more uniformly spread over each radius for later times. On the contrary, most of the lines are embedded in the interior of the WD at earlier times. 

\begin{figure}
    \centering    
    \includegraphics[width=0.485\textwidth]{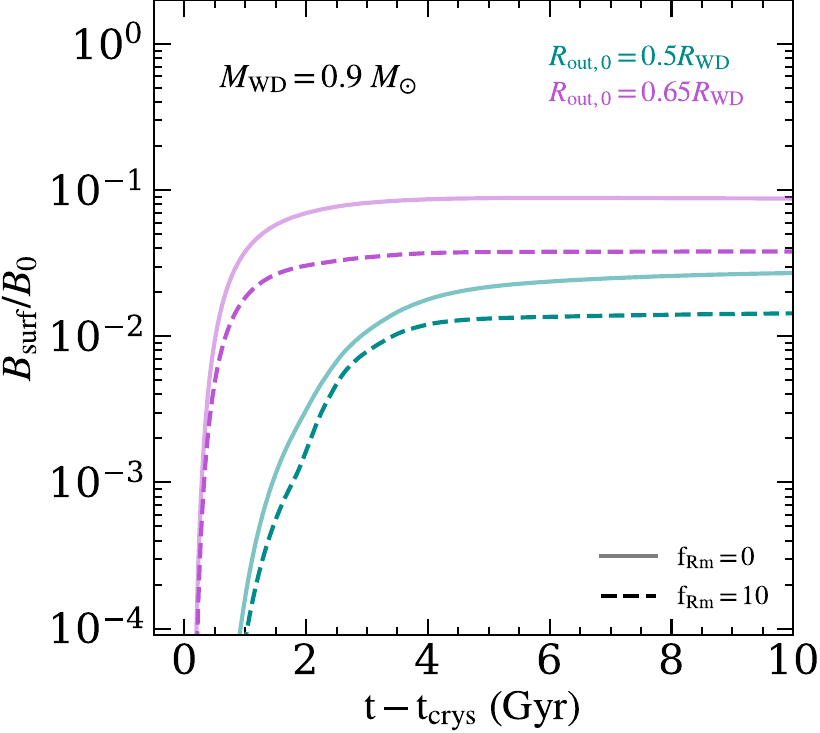}  
  \caption{Evolution over time of the surface magnetic field strength ($B_{\mathrm{surf}}$) for a $0.9\ M_{\odot}$ WD for different assumptions of the turbulent diffusivity factor $\mathrm{f_{Rm}}$ and the initial convection zone $R_{\mathrm{out,0}}$. After a couple of Gyr $B_{\mathrm{surf}}$ saturates to an almost constant value which is determined by the initial conditions $\mathrm{f_{Rm}}$ and $R_{\mathrm{out,0}}$.}
    \label{fig:B_time}
\end{figure}

In Figure \ref{fig:B_time} we show the time evolution of the magnetic field strength in the surface ($B_{\mathrm{surf}}$) of a 0.9 $M_{\odot}$ WD. In this case, we took different assumptions for $\mathrm{f_{Rm}}$ and $R_{\mathrm{out,0}}$. We see that independent of the initial conditions of the model, the surface magnetic field saturates to an almost constant value after a couple of Gyr from its formation. This is a direct effect of the crystallization front getting to the outer layers of the WD, as the solid core grows the diffusivity decreases freezing in the magnetic field, which means that the field no longer evolves significantly on Gyr timescales. The saturation value is different for each assumption of $\mathrm{f_{Rm}}$ and $R_{\mathrm{out,0}}$, which means, that these conditions entirely determine the maximum surface value that can be predicted. We discuss this problem in further detail in the following sections.

\subsection{Comparison with observed magnetic field strengths}
We now compare the surface $B$ values obtained from our transport models with the values of the observed magnetic fields in a sample of white dwarfs in the range of 0.45 to 1.05 $M_{\odot}$. We first selected a sample of MWDs from the Montreal White Dwarf Database (\citealt{Dufour2017}) limited within 100 pc distance to avoid problems with the stellar parameter estimations \citep[see][]{Caron2023}. Furthermore, we added some MWD from the 20 pc volume-limited sample of \citet{BagnuloLandstreet2021, BagnuloLandstreet2022} that are not in the Montreal 100 pc sample. We end up with a sample of 107 MWDs, where 86 are in the Montreal sample and 21 are exclusively from the 20 pc sample.

\begin{figure}
    \centering    
    \includegraphics[width=0.485\textwidth]{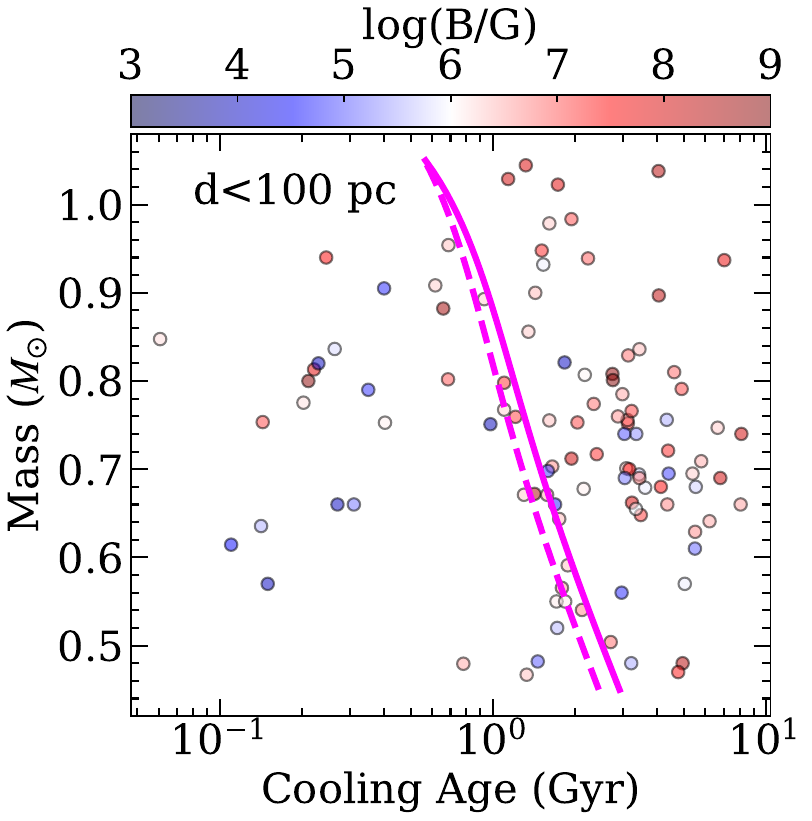}  
  \caption{Sample of magnetic white dwarfs within a distance of 100 pc and in the range of mass 0.45 to 1.05 $M_{\odot}$, extracted from the Montreal White Dwarf Database and the 20 pc volume-limited sample of \citet{BagnuloLandstreet2021, BagnuloLandstreet2022}. We show two lines that represent the expected age for the onset of crystallization of different WD masses for the cooling models from \citet{Bauer2023} described in Section \ref{sec:cooling_models} (solid magenta line), and for cooling models from \citet{BlatmanGinzburg2024} with oxygen-enriched cores obtained by increasing the $\mathrm{^{12}C(\alpha,\gamma)^{16}O}$ reaction rate during the stellar evolution (dashed magenta line, see text for more details).}
    \label{fig:MWD}
\end{figure}

In Figure \ref{fig:MWD}, we show the mass and cooling age of our sample of MWDs. The intensity of each object's surface magnetic field is presented using a color map. We added a magenta (solid) line that indicates an estimation of the cooling age at which crystallization is expected to start ($t_\mathrm{crys}$) for each stellar mass. For this, we made a polynomial fit between the mass and $t_{\mathrm{crys}}$ obtained from MESA models. Thus, we were able to estimate the crystallization time for WDs in the range of 0.45 to 1.05 $M_{\odot}$. With this threshold, we find 68 of the MWDs in our sample to be crystallizing, 54 from the Montreal sample, and 14 from the 20 pc sample.

Differences in the internal composition can lead to different cooling rates and initial crystallization times in WDs. Specifically, the reaction $\mathrm{^{12}C(\alpha,\gamma)^{16}O}$ significantly affects the initial composition profile of C/O white dwarfs, however, the rate of this $\alpha$ capture is poorly constrained \citep{deBoer2017, Chidester2022}. Recently, \citet{BlatmanGinzburg2024} investigated the effect of the $\mathrm{^{12}C(\alpha,\gamma)^{16}O}$ rate on $t_{\mathrm{crys}}$ and the expected time at which the crystallization-driven dynamo magnetic field would emerge from the surface of the WD. We use their MESA models that are in the $+3\sigma$ uncertainty range of the $\mathrm{^{12}C(\alpha,\gamma)^{16}O}$ reaction rate distribution from \citet{deBoer2017} and \citet{Mehta2022}, to compute $t_{\mathrm{crys}}$ for each mass as shown in the dashed magenta line in Figure \ref{fig:MWD}. The $+3\sigma$ uncertainty range creates composition profiles that are enriched in oxygen and crystallization starts earlier. Using this threshold we find 76 MWDs crystallizing in our sample, where 62 are from the Montreal sample and 14 from the 20 pc sample. 

\begin{figure*}
    \centering    
    \includegraphics[width=1.0\textwidth]{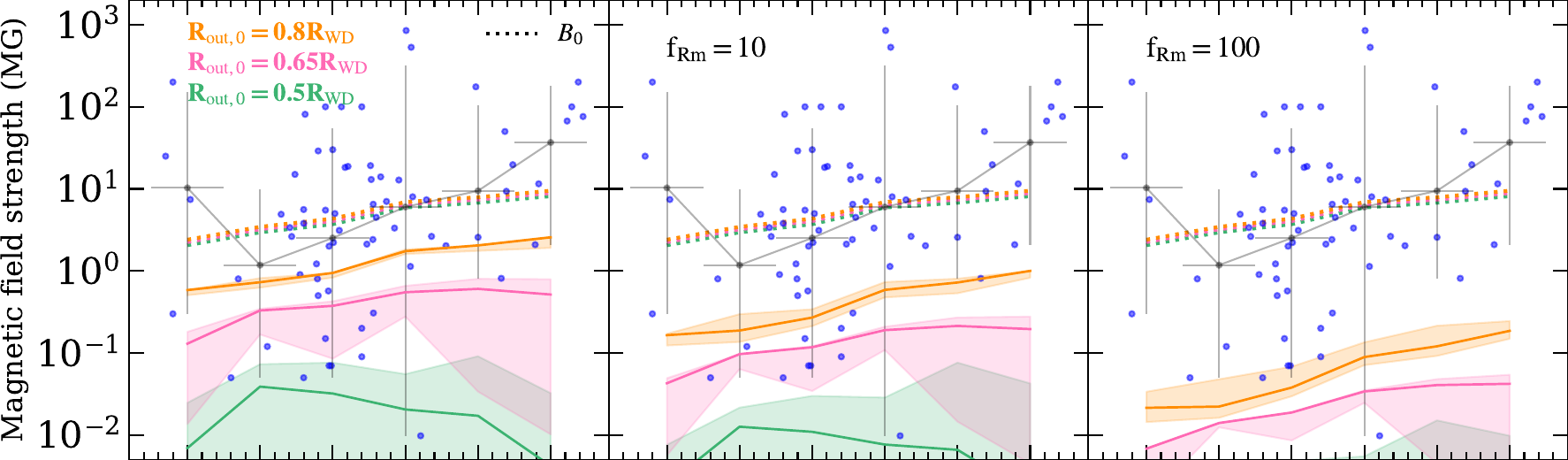} 
    \includegraphics[width=1.0\textwidth]{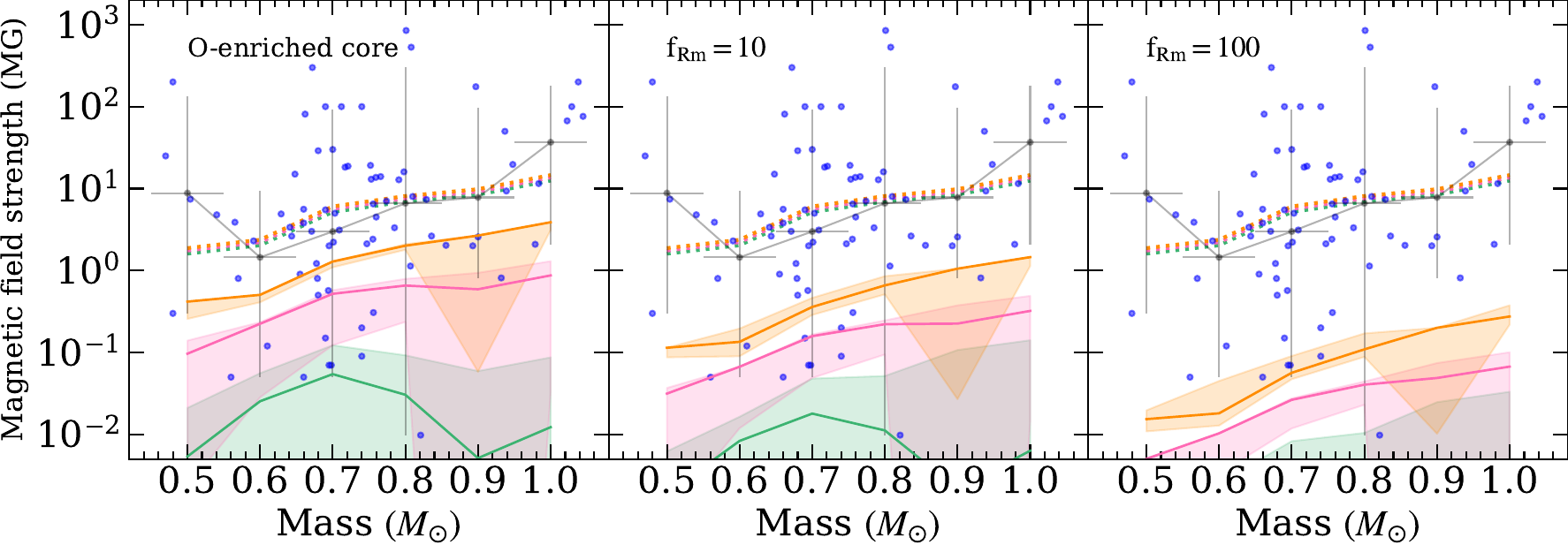} 
  \caption{Predicted surface magnetic field strength (solid lines and filled regions) from our transport models as a function of the WD mass for different assumptions of the initial size of the convection zone $R_{\mathrm{out,0}}$ and turbulent magnetic diffusivity factor $\mathrm{f_{Rm}}$. We compare these models with observed fields in MWDs that are expected to be crystallizing in our sample shown in Figure \ref{fig:MWD} for both assumptions of the onset of crystallization: top panels show the crystallizing MWDs obtained using the fiducial models from \citet{Bauer2023}, while the bottom panels are those obtained using the oxygen-enriched core models of \citet{BlatmanGinzburg2024}. We binned by mass the observations (black points with error bars) to set the cooling age for our evolution models, then the solid lines represent the final surface field when reaching the average cooling age in each bin, and the filled regions are the fields obtained for the one-standard deviation range of the cooling age (see text for details). The dashed lines represent the prediction for the initial magnetic field as discussed in Section 2.2.}
    \label{fig:Bobs_model1}
\end{figure*}

We took the WDs that are expected to be crystallizing from both cooling age limits we previously described and binned the data in mass intervals of 0.1 $M_{\odot}$ difference. Then each bin covers a range $M_{k\mathrm{-bin}}=(0.1\left[k,(k+1)\right]+0.45)M_{\odot}$ for $k=0,1,2,3,4,5$. To compare the surface magnetic fields from our models with the observed values, we obtained the average cooling age of the mass bins to set the time for the magnetic field evolution. 

In Figure \ref{fig:Bobs_model1}, we show the mass and observed magnetic field for our selected MWDs (blue points) considering $t_{\mathrm{crys}}$ from the models of \citet{Bauer2023} (top panels) and the O-enriched core cases (bottom panels), and compare them with the predicted surface magnetic field obtained from our transport models for different assumed values of $\mathrm{f_{Rm}}$ and $R_\mathrm{out,0}$. The mass bins used to obtain the cooling ages are presented with solid black points with error bars, where we also show the standard deviation of the average observed fields in each bin. The solid color lines of the transport models show the $B_{\mathrm{surf}}$ obtained for the average cooling age in each mass bin, and the filled regions enclose the values obtained within one standard deviation of the cooling age of the mass bins. Note that for a few cases, evolving the field for longer times does not produce larger $B_{\mathrm{surf}}$ values. This indicates that the surface field has reached a maximum and, as shown in Figure \ref{fig:B_time}, it slowly evolves stabilizing at a nearly constant value after a few Gyr.

\begin{figure*}
    \centering    
    \includegraphics[width=1.0\textwidth]{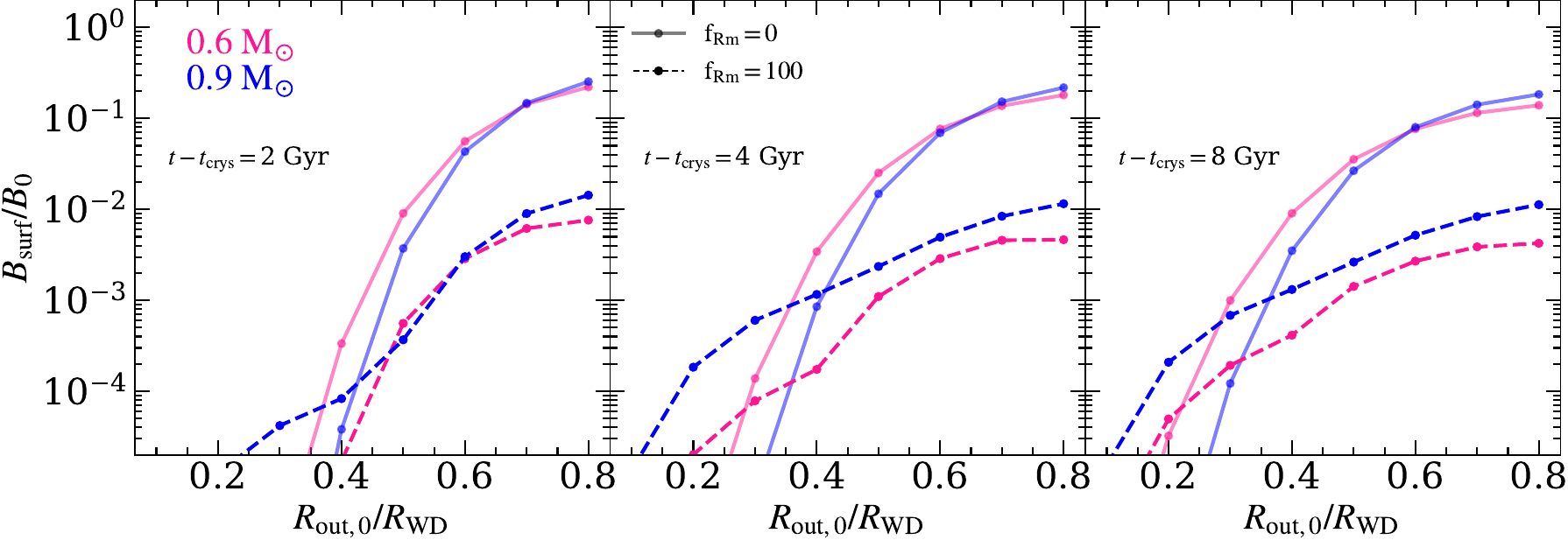} 
  \caption{Predicted surface magnetic field strength $B_\mathrm{surf}$ from our transport models, normalized by the initial field $B_{0}$, as a function of the initial size of the convection zone $R_{\mathrm{surf,0}}$, for two different masses, and for two assumptions on the turbulent diffusivity. We show $B_\mathrm{surf}$ at three different time evolution from the onset of the crystallization. We see that for smaller $R_{\mathrm{surf,0}}$ the predicted $B_\mathrm{surf}$ can be larger in certain cases when considering turbulent diffusivity, whereas the opposite behaviour occurs for larger $R_{\mathrm{surf,0}}$.}
    \label{fig:Rout}
\end{figure*}

The initial $B$ used in the diffusion models are the asymptotic values found using our analytical estimations for the stellar parameters towards the center of the WD as shown in Figure \ref{fig:B_vc} for each case of mass and $R_{\mathrm{out,0}}$. However, the efficient convection regime continues until $\tau$ decreases below 1. We find that the intensity of the magnetic field is reduced by about one order of magnitude when $\tau\gtrsim 1$. Thus, the predicted surface magnetic fields shown by our models are an upper limit (most optimistic case) for the values that we can compare with the observations.

From Figure \ref{fig:Bobs_model1}, we see that to match some of the observed magnetic fields in MWDs the initial size of the convection zone must be at least $0.5R_{\mathrm{WD}}$. A smaller $R_{\mathrm{out,0}}$ or a larger $\mathrm{f_{Rm}}$ significantly reduce the predicted $B_{\mathrm{surf}}$. The first effect occurs because the magnetic field cannot be rapidly transported to the surface since it is initially embedded in the inner layers of the WD, and it is mostly maintained inside as the central parts of the star solidify, increasing the conductivity. The second effect is due to the enhanced diffusivity of the magnetic field, which helps to transport the magnetic field, but at the same time, more magnetic energy is lost as the magnetic field diffuses to the surface.

 While for the values of $R_{\mathrm{out,0}}$ shown in Figure \ref{fig:Bobs_model1} the main effect of the turbulent diffusivity is to reduce $B_{\mathrm{surf}}$, this can be different for smaller $R_{\mathrm{out,0}}$. In Figure \ref{fig:Rout} we show the ratio of the surface magnetic field to $B_{0}$ as a function of $R_{\mathrm{out,0}}$ for two different masses and $\mathrm{f_{Rm}}$, and at different times. For the $0.9\ M_{\odot}$ model, we see that for $R_{\mathrm{out,0}}=$ 0.1 to 0.3 $R_{\mathrm{WD}}$ the turbulent diffusivity helps to transport the magnetic field to the surface quicker without significant loss of magnetic field strength, showing a larger $B_{\mathrm{surf}}$ prediction than in the case with only Ohmic diffusion at the three different times. For the $0.6\ M_{\odot}$ case, this behavior occurs only for $R_{\mathrm{out,0}}=$ 0.1 and 0.2 $R_{\mathrm{WD}}$. Except for these cases, turbulent diffusivity results in a reduced $B_\mathrm{surf}$.
 
Furthermore, in Figure \ref{fig:Rout} we observe that, overall, the more massive case shows larger $B_\mathrm{surf}/B_0$ than the less massive one when $\mathrm{f_{Rm}}>0$, whereas not including turbulent diffusion produces larger $B_{\mathrm{surf}}/B_{0}$ values for the less massive case. These features are due to the faster cooling and crystallization in more massive WDs, i.e.,  the larger conductivities in the solid phase mean that the magnetic field is frozen in at a larger value but also transported more slowly by Ohmic diffusion for the $0.9\ M_{\odot}$ case. 
Giving the predicted $B_{\mathrm{surf}}$ values in Figure \ref{fig:Rout}, some more massive WDs with magnetic fields of a few kG can be explained with the crystallization-driven dynamo if the initial convection zone is small and turbulent diffusivity dominates the transport of the field during the thermohaline convection regime.

\section{Summary and discussion}\label{sec:summary_conclusions}

We have studied the feasibility of the crystallization-driven dynamo to explain the observed magnetic fields in crystallizing carbon-oxygen white dwarfs with masses 0.45 to 1.05 $M_{\odot}$. To achieve this, we have extended the estimates of convective velocities and dynamo-generated magnetic field of \citet{Castro-Tapia2024} and \citet{Fuentes2024} to early times following crystallization, and then solved the induction (diffusion) equation to follow the evolution of the internal magnetic field after its formation. This allows us to estimate the strength of the surface magnetic field in the star at different cooling ages. In our calculations, we use fiducial white dwarf cooling models from MESA to obtain the necessary internal stellar parameters. We find surface fields that are generally too weak compared to observed magnetic fields. We therefore conclude that the crystallization dynamo can explain only a limited number of magnetic non-ultra massive C/O white dwarfs ($\lesssim 1.05\ M_{\odot}$), and some other mechanism is needed that can generate stronger surface magnetic fields to explain the remaining observed objects.

In general, we observe that the surface field shows an initial increase with time before saturating at an almost constant final value (Fig. \ref{fig:B_time}). This is because the electrical conductivity significantly increases in the solid phase, reducing the diffusivity, and freezing in the magnetic field. Then, the final surface magnetic field depends on how quickly the magnetic field can get to the outer layers of the star and the crystallization rate. If the transport of the field is too quick, more magnetic energy is diffused through the surface before the solid phase can stabilize its magnitude, on the contrary, if the crystallization is much faster than the transport of the field, most of the magnetic energy remains embedded in the interior of the white dwarf. 

We find that during the efficient regime of convection, lasting a few Myr after the onset of crystallization, the convective velocity and the induced magnetic field reach an asymptotic value, which sets a maximum strength of the magnetic field that can be generated (equation \ref{eq:Bmax}). This occurs due to the dramatic reduction of the local gravity considered in the buoyancy force for a small solid core, which cancels the divergence in the composition flux of light elements $\tau$ during these early times. Our predictions for the surface magnetic fields are optimistic in the sense that we used this maximum magnetic field as an initial condition for the evolution. If instead, we take the dynamo field at the time of transition between efficient and thermohaline convection, $B_\mathrm{surf}$ would be reduced by up to an order of magnitude. Further studies are needed to assess how the magnetic field responds to the changing composition flux during the dynamo phase. One caveat is that our estimates for the magnetic field strength generated in the dynamo rely on scaling arguments derived for planetary cores \citep{Davidson2013}; numerical simulations are needed to confirm these predictions for compositionally-driven convection in white dwarf conditions.

By comparing the predicted surface magnetic field values from our models with the observations of magnetic white dwarfs, we see that the initial convection zone and magnetic field must fill $\gtrsim 0.5$ to 0.8 of the radius of the white dwarf to match some observed fields after being transported for a few Gyr in the range of the cooling ages of the observed magnetic white dwarfs. This range for $R_{\mathrm{out,0}}$ is acceptable if we suppose that the initial convection zone fills the initial flat C/O composition profile in the core (without composition barriers) of fiducial white dwarf models for masses $\sim 0.5$-$1\ M_{\odot}$ \citep[see Figure 8 of][]{Bauer2023}. However, as shown in \citet{Fuentes2024}, including thermal diffusion in the criterion for the mixing may affect the extent of the convection zone during the efficient convection regime making it $\lesssim 0.1\ R_\mathrm{WD}$ before penetrating the outer layers. Since the convective boundaries cannot be correctly assessed given the current one-dimensional models for the fluid mixing in crystallizing white dwarfs, this problem demands more multidimensional simulations in this regime of compositionally-driven convection.

In our calculations, we assumed a dipolar magnetic field geometry. However, many observed MWDs have a more complex geometry.
For instance, \citet{Hardy2023} found that an off-centered dipole was necessary to fit spectroscopic observations of many MWDs. Thus, in the future, it is worth investigating if the crystallization-driven dynamo can generate off-centered dipole magnetic fields. This would modify the transport of the field, reducing, for example, the timescale for the Ohmic diffusion for the higher $l$ components \citep[see e.g., Table 1 of][]{Cumming2002}. The dipole offset would therefore likely be time-dependent.

We find that turbulent diffusion mostly reduces the predicted surface magnetic fields for $R_{\mathrm{out,0}}\gtrsim 0.5R_{\mathrm{WD}}$, so an enhanced spreading of the field dominates the more rapid transport to the surface, where $B_{\mathrm{surf}}$ quickly decays. However, for smaller values of $R_{\mathrm{out,0}}\lesssim 0.3R_{\mathrm{WD}}$, the more rapid transport does enhance $B_\mathrm{surf}$, especially for more massive white dwarfs. Still, the values obtained in this latter case are about $10^{-3}$ times smaller than the initial magnetic field generated during the dynamo, so that only magnetic fields of order $\lesssim$~kG can be achieved. In the future, evaluating the interaction between the magnetic fields and the convective motions in MHD simulations would help to better constrain this turbulent transport. Some recent studies have already given a first insight into the interaction of magnetic fields and thermohaline convection \citep[e.g.][]{Harrington_Garaud_2019, Fraser2024}, however, only the enhancement of the fluid mixing due to the magnetic field has been evaluated.


Our results suggest that the number of magnetic white dwarfs that can be explained by the crystallization-driven dynamo is limited. Once the diffusion of the magnetic field is taken into account, the surface values are reduced by at least a factor of $\sim0.2$ to 0.3 from the initial value. We find that we can only explain objects that are weaker than about 4 MG, being $\lesssim 1/3$ of the magnetic white dwarfs in our sample with measurements between $10^{4}$ to $10^{9}$ G. Thus, most magnetic white dwarfs still require an additional scenario to explain their origin. While observations suggest that stronger fields for white dwarfs with masses 0.5 to 1 $M_{\odot}$ are more frequent after the onset of crystallization \citep[][]{BagnuloLandstreet2022}, the hypothesis of a fossil magnetic field from previous stellar evolution stages cannot be discarded \citep[][]{Ferrario2015}. 

We also see that considering models with a more oxygen-enriched core, which leads to an earlier crystallization time, does not significantly change the predicted surface magnetic fields and the number of observations that we can explain. For this, we used WD profiles predicted from an enhanced $\mathrm{^{12}C(\alpha,\gamma)^{16}O}$ reaction during the helium-burning phase. However, even larger fractions of oxygen in the core have been inferred from asteroseismology measurements. Such variations in the chemical profile can be related to mixing processes like overshooting during the horizontal branch \citep{Giammichele2018, Giammichele2022}. Exploring these sources of uncertainty may be interesting to consider in the future and in particular the impact on the time of the onset of crystallization at each mass.

We have shown the limitations of the crystallization-driven dynamo theory, but we did not explore a scenario where the induced magnetic field and the compositional convection from this theory could amplify or modify a fossil field from stellar evolution. \citet{BlatmanGinzburg2024} discussed the idea that a fossil magnetic field (formed during the helium-burning phase) confined deep in the C/O core can be transported to the surface by crystallization-driven convection. Our results in Figure \ref{fig:Rout} show that if the fossil field is confined to $<0.5\ R_{\mathrm{WD}}$, the strength of the surface field once it emerges is reduced by at least a factor of 100 compared to its initial value, and potentially a factor of $10^4$. Further work describing the evolution of the magnetic field formed by different scenarios is needed to fully assess the origin of most magnetic white dwarfs.

\begin{acknowledgements}
 We thank Sivan Ginzburg for very useful comments on our previous work, and Mike Montgomery, Bart Dunlap, and Sihao Cheng for their questions and comments related to the implications of gravity for the small core size at the onset of crystallization. A.C. acknowledges support by NSERC Discovery Grant RGPIN-2023-03620. A.C. and M.C.-T. are members of the Centre de Recherche en Astrophysique du Québec (CRAQ) and the Institut Trottier de recherche sur les exoplanètes (iREx). A.C.~is grateful to the Isaac Newton Institute for Mathematical Sciences, Cambridge, for support and hospitality during the programme ``Anti-diffusive Dynamics: from sub-cellular to astrophysical scales'' supported by EPSRC grant no EP/R014604/1.
\end{acknowledgements}

\bibliographystyle{aasjournal}

\end{document}